\documentclass[pre,aps,preprint,showpacs]{revtex4} 
\usepackage{revsymb,amsmath}
\usepackage{graphicx}
\begin{document}
\author{Steffen Trimper, Knud Zabrocki}
\affiliation{Fachbereich Physik, Martin-Luther-Universit\"at,D-06099 Halle Germany}
\email{trimper@physik.uni-halle.de}
\author{Michael Schulz}
\affiliation{Abteilung Theoretische Physik, Universit\"at Ulm\\D-89069 Ulm Germany}
\email{michael.schulz@physik.uni-ulm.de}
\title{Memory-Controlled Diffusion}
%\draft
\date{\today }

\begin{abstract}

Memory effects require for their incorporation into random-walk models an extension of the conventional 
equations. The linear Fokker-Planck equation for the probability density $p(\vec r, t)$ is generalized 
to include non-linear and non-local spatial-temporal memory effects. The realization of the memory kernels 
are restricted due the conservation of the basic quantity $p$. A general criteria is given for the existence 
of stationary solutions. In case the memory kernel depends on $p$ polynomially the transport is prevented. 
Owing to the delay effects a finite amount of particles remains localized and the further transport is 
terminated. For diffusion with non-linear memory effects we find an exact solution in the long-time limit. 
Although the mean square displacement shows diffusive behavior, higher order cumulants exhibits differences 
to diffusion and they depend on the memory strength. 
\end{abstract}
\maketitle

\section{Introduction}

Although the crucial factors governing the dynamics of systems, comprising many units, consist  
of interaction and competition, there is an increasing interest to include memory effects as a further 
unifying feature of complex physical \cite{6a,6b} as well as biological systems \cite{6c}. Recently \cite{fk} 
memory effects in correlated anisotropic diffusion are studied in nanoporous crystalline solids. Likewise 
the effects of transport memory are discussed in Fisher`s equation \cite{ah}, also applicable for bacterial 
population dynamics \cite{kk}, resulting in non-linear damping and traveling wave solutions \cite{abk}. The 
transport with memory, depending on the survivability of a population, is analyzed in \cite{h}.    
In the present paper we emphasize in an analytical solvable model, that the generic behavior of the system may be 
changed when additional non-linear delay effects are included into the consideration. In particular, we discuss 
the transport behavior which is realized after a sufficient accumulation time and after cumulating particles 
within a spatial region. Thus the transport processes are affected by additional spatial correlations. The simplest 
way to describe transport in a medium is based on a random walk. The probability $p(\vec r,t)$ to find a particle 
at the position $\vec r$ at time time $t$ is governed by the Fokker-Planck equation \cite{ga}
\begin{equation}
\partial_tp(\vec r, t) = \mathcal{M}(\vec r, t; p, \nabla p)\,.  
\label{1} 
\end{equation}
Here the operator $\mathcal{M}$ consists of an diffusive and a driving part. In case the diffusive part is 
relevant we follow scaling arguments, extended also to more general processes in \cite{cd}. Based on that one 
can identify  a diffusive time scale given by $\tau_D \simeq l^2/D $. Here 
$l$ is a typical length scale, where the probability is changed significantly. 
In case the transport process is realized after a spatial-temporal accumulation processes the time evolution 
of the probability could also depend on the history of the sample to which 
it belongs, i.e. the changing rate of the probability should be influenced by the changing rate in the past. 
Thus the evolution of the probability $p(\vec r, t)$ has to be supplemented by memory terms. Such a term models, 
for example the way on which a seed probability at a certain time $t'$ had been accumulated by a 
delayed transport mechanism, originated by the surrounded environment of the particle. In general, the changing 
rate of $p$ at time $t$ is also determined by the accumulation rate at a former time 
$t' < t$. In between, i.e. within the interval $\tau = t - t'$, the particles are enriched while 
changing the probability at $t'$. Regardless that process the available amount of particles at time $t$ is 
governed by an instantaneous transport term as well as by the changing rate at former times $t'$. 
Consequently the evolution Eq.~(\ref{1}) should be modified according to  
\begin{equation}
\partial _{t}p(\vec r,t) = \mathcal{M}(\vec r,t; p, \nabla p)  
+\int\limits_{0}^{t}dt'\int\limits_{-\infty }^{\infty }d^dr^{\prime}
\mathcal{K}(\vec r- \vec r^{\prime},t-t'; p, \nabla p)\mathcal{L}(\vec r^{\prime}, t'; p,\nabla p)
\label{2}
\end{equation}
This equation is of convolution type. Whereas the operator $\mathcal{M}$ characterizes the instantaneous and 
local processes, both operators $\mathcal{K}$ and $\mathcal{L}$ are responsible for the delayed processes. In general 
the operators $\mathcal{M}, \mathcal{K}$ and $\mathcal{L}$ are assumed to be non-linear in $p(\vec r, t)$ and 
$\nabla p(\vec r,t)$. They have to be specified according to the physical situation in mind. In particular 
we show that the operators are restricted when $p(\vec r,t)$ is a conserved quantity.\\ 
It is well-known that evolution equations with such a kind of memory kernels 
can be derived following the well established projector formalism due to \cite{mori}, see also \cite{naka}. 
Notice that Eq.~(\ref{2}) is an effective single particle equation embedded in a $N$-particle system. The main 
approach for the further analysis consists of assuming that the time scale of the memory effects is controlled 
by the time scale of the probability itself. The environment of a single particle is determined by the transport 
processes of the remaining $N$ particle which follows approximately the same underlying physical processes. 
In the next section the operators $\mathcal{K}$ and $\mathcal{L}$ are specified based on that argument.\\              
Our model can be grouped into the increasing efforts to discuss delay and feedback mechanism. The analysis of 
the projector formalism \cite{mori} had been successfully applied for the density-density correlation function 
studying the freezing processes in under-cooled liquids \cite{l,goe}. Recently a Fokker-Planck equation with a 
non-linear memory term was used to discuss anomalous diffusion in disordered systems \cite{ss}. The results 
could be confirmed by numerical simulations including diffusion on fractals \cite{bst}, 
see also \cite{t,mo}. Moreover, it was argued \cite{loc} that mobile particles 
remain localized due to the feedback-coupling. Notice that a formal solution of 
whole class of non-Markovian Fokker-Planck equations can be expressed through the solution 
of the Markovian equation with the same Fokker-Planck operator \cite{s}. The non-Gaussian 
fluctuations of the asset price can be also traced back to memory effects \cite{st}. 
An additional cumulative feedback coupling within the Lotka-Volterra model, which may stem 
from mutations of the species or a climate changing, leads to a significant different 
behavior \cite{zab} compared to the conventional model. If the Ginzburg-Landau model 
for the time evolution of an order parameter is supplemented by a competing memory term,  
the asymptotic behavior and the phase diagram is completely dominated by such a term 
\cite{zab1}. Whereas the feature of the approach, proposed in those papers, consists of self-organization, 
i.e. the time scale of the memory is determined by the relevant variable itself, for instance the 
concentration, there is a broad class of models with external delay effects \cite{oy,ge,fe}, for a survey 
and applications in biology see \cite{mur}. That case is characterized by a given external memory 
kernel. The differences of both approaches will be discussed also in \cite{zab2}. The spreading of an agent 
in a medium with long-time memory, which can model epidemics, is 
studied in \cite{gr}. Time-delayed feedback control is an efficient method for stabilizing 
unstable periodic orbits of chaotic systems \cite{p} where the time delay may induce various 
patterns including traveling rolls, spirals and other patterns \cite{j}. 
The influence of a global feedback is studied recently in a bistable system \cite{sa}, where the   
purpose of that paper is a discussion of the domain-size control by a feedback.\\ 
In view of the large variety of systems with feedback couplings it seems to be worth to study 
simple models, which still conserve the crucial dynamical features of evolution models such as 
inherent non-linearities and moreover, as a new ingredient, delayed feedback-couplings. In the 
present paper we discuss the influence of a non-Markovian memory term on transport processes.  
The retardation effects are characterized by the memory kernel $\mathcal{K}$, which is chosen in such a manner, 
that it competes with the conventional diffusion part and obeys the same symmetry properties. In particular, 
we discuss the transport behavior under the inclusion of non-linear memory term. Under that conditions the 
characteristic time $\tau _D$ is modified. Accordingly we demonstrate, the long-time behavior is dominated 
apparently by the delay effects. The system is able to reach a stationary state with a finite probability.

\section{Model} 

In this section we specify the model, defined by Eq.~(\ref{2}), under the assumption that basic 
quantity $p(\vec r, t)$ is conserved. This condition is realized by 
\begin{equation}
\dot{P}(t) = \frac{d}{dt} \int_{-\infty}^{+\infty} d^{d}r p(\vec r, t) = 0\quad .
\label{ev1}
\end{equation} 
To preserve $p$ the instantaneous term $\mathcal{M}$ has to be related to a current, e.g. 
$\mathcal{M}\propto \nabla \cdot \vec j $. Choosing natural boundary conditions $p=0$ at the boundary), we get 
after a Laplace transformation
\begin{equation}
z P(z) - P_0 = \hat{K}(z) \hat{L}(z) \quad\mbox{with}\quad \hat{K}(z) = \int d^dr\, \mathcal{K}(\vec r, z),\qquad 
\hat{L}(z) = \int d^dr\, \mathcal{L}(\vec r, z)\,,
\label{ev2}
\end{equation}  
with $P_0 = P(t=0)$. For an arbitrary polynomial kernel $\hat{K}$ the conservation law is in general not 
fulfilled provided the operator $\mathcal{L}$ is simply defined by $\mathcal{L} \equiv - \partial_t p(\vec r, t)$ 
(the minus sign is only for convention). Making this ansatz we conclude from Eq.~(\ref{ev2}) 
$$
[z P(z) - P_0][\hat{K}(z) + 1] = 0
$$
and consequently the conservation law is guaranteed. Physically, the assumption for $\mathcal{L}$ means, that we take into account a coupling of the rates, e.g. the evolution at the observation time $t$ is directly coupled to the changing rate at $t' < t$. Processes in the past will be permanently reevaluated at present time. In doing so the memory kernel gives rise to a coupling of the time scales. In the vicinity of the upper limit $t^{\prime} \simeq t$ the memory term reads 
$\mathcal{K}[\vec r, 0,\,p(\vec r, 0)]\partial_t p(\vec r, t)$, i.e. a momentary change at the observation time $t$ is coupled to the value at the initial time $t = 0$. Therefore the very past is related to the instantaneous value of 
$p(\vec r, t)$. In the opposite case, at the 
lower limit $t^{\prime} \simeq 0$, the change of the quantity $p(\vec r, t)$ near to the initial value 
$\partial_{t^{\prime}}\,p(\vec r, t^{\prime} = 0)$ is directly coupled to the instantaneous value $p(\vec r, t)$ via the kernel. In such a manner the memory part represents a weighted coupling of the behavior at the initial time and the observation time. Due to the coupling of the rates the long-time behavior of the system will be modified. 
One reason for that could be that the reacting species are embedded into an environment of all 
the other particles of the system. Due to the mutual interaction, reactants, lacking at time $t$, were 
annihilated at an earlier time $t^{\prime}$. Especially in sufficiently complex diffusion-reaction systems the feedback 
and memory effects should be relevant. In such system additional degrees of freedom like in flexible macromolecules 
in melts or in concentrated solutions \cite{dg}, nematic elastomer \cite{af} or in biology \cite{mur}. 
In that context one is interested in the description of effluent reprocessing plants in systems with closed 
water circulation. A special ecosystem of aerobic and anaerobic microorganisms is evolved in the 
clarifiers of such systems due to natural immigration or due to additional allowance. The living conditions 
of the microorganisms are mutually associated via the exchange of intermediate catabolic products. 
Each change of the concentration of one species will be stored in the food chain and 
effects the evolution of this species at a later time. Furthermore,
the partial mixing in the clarifiers by convection and diffusion processes
enlarges the effects over the whole system, so that the memory integral
introduced in Eq.~(\ref{1}) includes both the time and the spatial coordinates. 
This special example may be extended also to other complex biological, chemical
or engineering problems with various hidden degrees of freedom, which are able to 
influence the evolution of a selected component significantly, for instance 
by biological interaction with other species via the food chain or via biological competition. 
Such effects which are partially observable, could contribute to the memory term. \\
\noindent From Eq.~(\ref{2}) we can make some general statements for an arbitrary kernel 
$\mathcal{K}(\vec r, t; p, \nabla p)$ and $\mathcal{L}(\vec r, t; p, \nabla p) = - \partial_t p(\vec r, t)$. 
As stressed before the last condition  
guarantees the conservation of $p(\vec r, t)$. After Fourier transformation with respect to the spatial coordinate and Laplace transformation with respect to the time, 
we get from Eq.~(\ref{2})
\begin{equation}
p(\vec k, z) = \frac{p_0(\vec k)}{z + \hat{D}(\vec{k}, z)\,k^2}\quad\text{with}\quad 
\hat{D}(\vec{k}, z) = \frac{D}{1 + \mathcal{K}({\vec{k}}, z)}\,,
\label{ev10}
\end{equation} 
where $\mathcal{K}(\vec k, z)$ is the Fourier-Laplace transformed kernel. Assuming a regular behavior of the 
kernel we make the ansatz
\begin{equation}
\mathcal{K}(\vec k, z) = \frac{B(\vec k)}{z} + \Omega (\vec k, z)\quad\text{with}\quad\lim_{z \to 0}\Omega({\vec k}, z)= \text{finite}\quad .
\label{ev11}
\end{equation}
Provided the kernel reveals a finite stationary value $\lim\limits_{t \to \infty} \mathcal{K}(\vec k, t) \equiv B(\vec k)$, then 
$p(\vec r, t)$ yields a stationary solution, too. Inserting Eq.~(\ref{ev11}) in Eq.~(\ref{ev10}) we obtain
\begin{align}
p(\vec k, z) &= \frac{g(\vec k)}{z} + \psi (\vec k, z)\nonumber\\
\text{with} \quad g(\vec k) &= \frac{p_0(\vec k)}{B(\vec k) + D k^2}\quad\text{and}\quad
\psi  (\vec k, z) = \frac{p_0(\vec k)\,Dk^2\,[\,1 + \Omega (\vec k, z)]}{[\,A(\vec k) + D k^2)\,][\,z(1 + \Omega (\vec k, z)) + 
B(\vec k) + Dk^2]}\,. 
\label{ev12}
\end{align} 
Summarizing the results we conclude that the model, following Eq.~(\ref{2}) for the conserved quantity $p(\vec r, t)$,    
gives rise to a non-trivial stationary solution $g(\vec k$), or equivalent $g(\vec r)$, if the Laplace transformed 
kernel $\mathcal{K}(\vec k, z)$ satisfies the conditions
\begin{equation}
\lim_{z \to 0} z \mathcal{K}(\vec k, z) \neq 0\,.
\label{ev14a}
\end{equation}
This result is a generalization of a previous one obtained for a homogeneous system \cite{trizab}.\\
Since $p(\vec r, t)$ is a probability density we can derive an evolution equation for the mean square displacement 
$s(t) = <\vec r^{~2} >$. Provided the function $p(\vec r, t)$ is already normalized, one finds
\begin{equation}
\frac{d}{dt}s(t) = 2 d D - \int\limits_{0}^{t} dt' \hat{K}(t-t')\hat{L}_2(t') - \int\limits_0^t dt' \hat{K}_2(t-t') \hat{L}(t')\,,
\label{19}
\end{equation}
where $\hat{K}$ and $\hat{L}$ are defined in Eq.~(\ref{ev2}) and $\hat{K}_2$ and $\hat{L}_2$ are the 
corresponding second moments. In the simplest case without memory and $\mathcal{M} = D \nabla^2 p(\vec r, t)$ one  
concludes  from Eq.~(\ref{19}) $s(t) = 2 d D t$. In the next section we discuss two cases with non-zero kernel 
$\mathcal{K}$ depending on $p(\vec r, t)$ itself, e.g. the time-space scale of the memory is given by the scales 
where the variable $p(\vec r, t)$ becomes relevant.  

\section{Non-linear memory}
For applications we consider the case of diffusive instantaneous term given by 
\begin{equation}
\mathcal{M} = D \nabla ^2 p(\vec r, t)\,,
\label{inst}
\end{equation}
where $D$ is the diffusion constant. In the first subsection let us illustrate the influence of memory by specifying 
the kernel $\mathcal{K}$ by a power law in the quantity $p$. Hereby we follow the ideas discussed recently \cite{ss,loc}.
\subsection{Polynomial kernel}
Based on the analysis of under-cooled liquids in the frame of mode-coupling theory \cite{l,goe} the 
random walk had been analyzed in a glass-like environment \cite{ss}. In that case the memory kernel becomes
\begin{equation}
\mathcal{K}(\vec r, t) = \lambda p^2(\vec r, t)\,,
\label{mem}
\end{equation}
where $\lambda $ is the strength of the memory. Because in that case $\hat{K}(t)$ is different from zero and 
in general different from $-1$ one has to choose the operator $\mathcal{L} = - \partial_t p(\vec r, t)$ to preserve 
the basic quantity $p$. The evolution equation reads
\begin{equation}
\partial _{t}p(\vec r,t) = D \nabla ^2 p(\vec r, t) - \lambda \int\limits_{0}^{t}dt'\int\limits_{-\infty }^{\infty }d^dr'
p^2(\vec r - \vec r^{~\prime},t-t')\partial_{t'}(\vec  r^{~\prime}, t')\,.
\label{mem1}
\end{equation}
Making a simple scaling transformation $\vec r \to \Lambda \vec r,\,t \to \Lambda ^z t,\, p \to \Lambda ^{-d} p$, 
the evolution equation with the memory kernel, defined by Eq.~(\ref{mem}) remains invariant by the replacement 
$D \to D \Lambda ^{2-z}$ and $\lambda \to \lambda \Lambda ^{d-2}$. From here one gets the critical dimension 
$d_c = 2$. Defining according to Eq.~(\ref{ev11}) $\hat{K}(t) = \lambda \int d^dr\,p^2(\vec r, t)$, the mean square 
displacement $s(t)$ obeys as follows from Eq.~(\ref{19}) after Laplace transformation        
\begin{equation}
s(z) = \frac{s_0[z + z \hat{K}(z)] + 2 D d }{z\,[\,z + z \hat{K}(z)]}\quad\text{with}\quad s_0 = s(t=0)= 0
\label{l}
\end{equation}
As demonstrated in \cite{loc} the non-linear transport process with the kernel $\mathcal{K}$ in Eq.~(\ref{mem}) 
allows a stationary solution. Following Eq.~(\ref{ev11}) the stationary part of the kernel reads
$$
\hat{K}(z) = \frac{\tilde{B}}{z}\quad\text{with}\quad \tilde{B} \equiv B(\vec k = 0) = \lambda \int\limits_{-\infty}^{\infty}d^dr\,g^2(\vec r\,) \neq 0\,,
$$
where $g(\vec r)$ is the stationary solution according to Eq.~(\ref{ev12}). 
As a result the mean square displacement exhibits a finite stationary solution 
\begin{equation}
\lim_{t \to \infty} s(t) = \frac{2 d D}{\tilde{B}}\propto \frac{D}{\lambda }\,,
\label{fin}
\end{equation}
which is only reasonable for a positive memory strength $\lambda > 0$ in agreement with  \cite{ss,loc}. 
For simplicity the initial value $s_0$ is assumed to be zero. In the short time regime, characterized by 
$z \gg 0 $, we find $s(z)\simeq 2d D z^{-2}$. As a consequence the mean square displacement increases 
linearly in $t$, whereas $s(t)$ becomes constant in the long-time limit. The characteristic cross-over time from 
a linear increase in time to the constant stationary value, given by Eq.~(\ref{fin}), is given by $t_c = 1/\tilde{B}$. 
A linear stability analysis shows that the finite stationary solution is stable. Physically a  finite stationary limit can be identified with localization. Owing the self-organized memory effects a finite amount of particles remains localized preventing a further transport. The analytical results are well confirmed by numerical simulations \cite{bst}. 
Note that one can also calculate higher order moments like $w(t) = < (\vec r^{~2})^2>$ which offers also a stationary 
behavior which fulfills the relation 
$$
\lim_{t \to \infty} w(t) = \left(1+\frac{2}{d}\right)\,\left(2 - \frac{\tilde{B_2}}{D d}\right)\lim_{t \to \infty}s^2(t)
$$
with $\tilde{B}_2 = \lambda \int d^dr\,\vec r^{~2} g^2(\vec r\,)$. A similar behavior can be deduced for another power law memory kernel. 

\subsection{Non-linear diffusion}

Now we regard as a further realization of Eq.~(\ref{2}) with a conserved $p(\vec r, t)$ a non-linear 
diffusion equation. Here the non-linearity may be originated by the memory. For example, a Brownian particle moving 
in a strongly disordered medium, can be subjected to additional delay effects due to the interaction with the environment. Another application could be realized by transport of food in plants, where the food is temporally stored in certain 
cells and released later as required.\\ 
\noindent In terms of our model the instantaneous term is chosen as conventional diffusion, but the 
memory kernel $\mathcal{K}$ is also related to gradient terms, i.e. the $\mathcal{K} = \mu \nabla ^2 p$. This 
realization is a generalization of Cattaneo`s law \cite{ca,me,aha}, where the current is assumed to be 
$\vec j(\vec r, t + \tau) \propto \nabla p(\vec r, t)$. Thus, the model includes not only the instantaneous and local coupling to the gradient but also there appears a non-local, 
time-delayed diffusive mechanism. The evolution equation reads after partial integration  
\begin{equation}
\partial _{t}p(\vec r, t) = D \nabla^2 p(\vec r, t) - \mu \int\limits_{0}^{t}dt'
\int\limits_{-\infty }^{\infty }d^dr' \,
\nabla p(\vec r - \vec r^{~\prime}, t - t') \cdot \partial_{t'} \nabla p(\vec r^{~\prime},t')\quad . 
\label{ev14}
\end{equation}
The non-linear term is a competitive one to the conventional diffusion term. Both concentration fluctuations at 
times $t$ and $t'$, respectively contribute to the behavior of the system. Different to the diffusion constant 
$D > 0$, the sign of the parameter $\mu$ can be positive and negative indicating the "direction" of the feedback. 
Note that in this case the memory is not determined by an external function as in the previous section, but the 
memory is self-organized by the quantity $p(\vec r, t)$ and $\nabla p$. The non-linear term is of the same form 
as used in the so called KPZ-equation \cite{kpz}.\\ 
\noindent As discussed before, Eq.~(\ref{ev14}) remains invariant under a simple scaling transformation for 
the spatial and time variable as well as the probability density by replacing
$$
D \to \Lambda ^{2-z}\,D,\quad \mu \to \Lambda ^{2-z}\,\mu \,.
$$
From here we conclude that the non-linear feedback term in Eq.~(\ref{ev14}) behaves like diffusion. In terms of the effective diffusion parameter $\hat{D}$, introduced in Eq.~(\ref{ev10}), the model leads to 
$$
\hat{D}(\vec k, z) = \frac{D}{1 + \mu k^2 p(\vec k, z)}\,.
$$
For a positive memory parameter $\mu > 0$, it results $\hat{D} < D $. In the long wave length limit $\vec k = 0$ 
we find $\hat{D}(\vec k = 0, z) = D$, while in the long time limit corresponding to $z \to 0$, the effective 
diffusion parameter is different from $D$. According to Eq.~(\ref{ev12}) the model exhibits the two stationary solutions
$$
g_+(\vec r) = 0\quad \text{and}\quad g_-(\vec r) = p_0(\vec r) - \frac{1}{\kappa^2} \delta (\vec r)\,.
$$ 
Here we have introduced the scale-free parameter $\kappa ^2 = \mu/D$. 
The non-trivial solution is accessible if $g_- > 0$ which is fulfilled for $\mu > 0 $. Because both solutions are 
stable within a linear stability analysis, compare Eq.~\ref{ev12}), one has to study the complete dynamics. To that 
aim let us perform a combined Fourier and Laplace transformation of Eq.~(\ref{ev14}). The resulting quadratic equation 
offers two solutions $p_{\pm }(\vec k, z)$ with
\begin{align}
p_{+}(\vec k, z) &= \frac{p_0(\vec k)}{2A_1}\sum_{n=1}^{\infty} (-a_n)^{n+1}\frac{(4A_1)^n z^{n-1}}{(z + A_2)^{2n-1}}
\nonumber\\
p_-(\vec k, z) &= -\frac{p_0(\vec k)}{A_1}-\frac{A_2}{A_1 z} - p_+(\vec k, z) \nonumber\\
\text{with}\quad A_1(\vec k) &= \mu k^2 p_0(\vec k), \quad A_2(\vec k) = k^2\,[\,D - \mu p_0(\vec k)\,],\quad \text{and}
\quad a_n = \frac{\Gamma (n-\frac{1}{2})}{2 \sqrt{\pi} n!}
\label{ev15}
\end{align}
After a tedious but straightforward calculation we find an exact expression for $p_+(\vec k, t)$ where we have used for simplicity $p_0(\vec k) = p_0$. It results  
\begin{equation}
p_{+}(\vec k, t) = \frac{p_0 e^{-A_2 t}}{\sqrt{\pi }}\sum_{n=1}^{\infty}
\frac{\Gamma (n-1/2)(-4A_1t)^{n-1})}{n \Gamma ^2(n)}M(1-n,n;A_2t) \equiv f(k^2t)\,,
\label{ev16}
\end{equation}
where $M(a,b;x) \equiv _{1}F_1(a,b;x)$ is Kummer`s confluent hypergeometric function \cite{abram}. Using the 
expressions for the quantities $A_a$ and $A_2$, defined in Eq.~(\ref{ev15}), one concludes that the solution 
$p_+(\vec k, t)$ is a function of $k^2t$ as in case of conventional diffusion. Therefore, after Fourier transformation the solutions can be written as 
\begin{equation}
p_+(\vec r, t) = \frac{1}{t^{d/2}}F\left(\frac{\vec r^{~2}}{t}\right)\,.
\label{ev18}
\end{equation}
where the scaling function $F(w)$ is different from the diffusive case. To proceed further we use the asymptotic 
representation of Kummer`s function leading to 
\begin{equation}
p_+(\vec k, t) \simeq p_0 e^{-A_2 t} \frac{I_1(\,2t\sqrt{A_1\,A_2})}{t\sqrt{A_1 A_2}}\,,
\label{ev17}
\end{equation}
where $I_1(y)$ is the modified Bessel function \cite{abram}. In the long time limit the solution $p_+$, belonging to the 
stationary solution $g_+ = 0$, is stable whereas the other solution $p_-$, related to $g_-$ becomes unstable.
Actually from Eq.~(\ref{ev17}) we can get the spatiotemporal solution. Remark that an divergent factor, proportional to 
$k^2$ in the denominator of Eq.~(\ref{ev18} is compensated by the modified Bessel function $I_1$. This is also the 
reason that one can not use the asymptotic expansion of $I_1$ in estimation the scaling function $F(w)$ in 
Eq.~(\ref{ev18}). 
We find
\begin{equation}
F(y) = \frac{p_0 e^{-\frac{\vec r^{~2}}{4(D-\mu p_0)t}}}{(4\pi (D - \mu p_0) t)^{d/2}}
\sum_{n=0}^{\infty} \left(\frac{p_0 \mu }{d-\mu p_0}\right)^n \frac{(2n)!}{n! \Gamma (n+2)}
L_{2n}^{d/2-1}\left( \frac{\vec r^{~2}}{t}\right)\,,                    
\label{sca}
\end{equation}
with the Laguerre polynomials $L_a^b(x)$. To establish the difference to conventional diffusion let us study 
the mean square displacement $s(t) = <\vec r^{~2} >$. According to Eq.~(\ref{19}) $s(t)$ obeys the equation  
\begin{equation}
\frac{d}{dt}s(t) = 2 d D - \int_0^t dt' \hat{K}(t-t')\hat{L}_2(t') - \int_0^t dt' \hat{K}_2(t-t') \hat{L}(t')\,.
\label{20}
\end{equation}
Within the model, discussed in this section, we find from Eq.~(\ref{20}) 
$s(t) = 2 d D t$. The mean square displacement is identical with that for diffusion. It is independent on the memory strength $\mu $. The difference to diffusion becomes visible in higher order moments. Using the same procedure as for  deriving Eq.~(\ref{20}), we obtain for $w(t) = < (\vec r^{~2})^2>$ the equation
$$
\frac{d}{dt}w(t) = s(t) \left[4D(d+2) + 2 \mu (d + 4)\right]\,.
$$  
Whereas the fourth order cumulant $C_4(t) = w(t) - 3 s(t)$ for conventional diffusion is $C_4(t) = 2d(1-d)(2Dt)^2\,$, it results for the model with memory
$$
C_4(t) = (2Dt)^2\,2d\,\left[\,1-d + \frac{\mu }{4D}\,(d+4)\right]\,.
$$
The cumulant depends on the ratio $\mu /D = \kappa ^2$. The result can be extended to higher order cumulants.

\section{Conclusions}

In this paper we have extended the conventional modeling of diffusive processes by including non-Markovian 
memory terms within the evolution equation. The additional terms are chosen in such a manner that the 
relevant variable $p(\vec r, t)$ can be normalized. By this requirement the form of the memory term is restricted 
to a class where the changing rate of $p$ at the observation time $t$ is coupled to the changing rate of $p$ 
at a former time $t'$. Insofar the delay effect offers a long time memory due to $0 \leq t' \leq t$. Further 
the model exhibits also a long range memory because the kernel $\mathcal{K}$ depends on $\vec r - \vec r^{~\prime}$. 
As a new ingredient the memory term is determined by the basic quantity $p(\vec r, t)$ itself, i.e. the 
memory is dominated by the spatial-temporal scale of the probability density $p$. In the paper we demonstrate that 
such a self-organized feedback coupling may change the dynamical behavior of the system essentially, in particular 
due to the non-linearity of the memory effects. In case the memory offers a power law dependence on $p$ with a 
memory strength $\lambda $ the particle performing a random walk, can be localized and therefore a further transport 
is prevented. This situation is realized for an "attractive" memory strength $\lambda > 0$, where the return probability dominates the dynamical behavior. There exists a finite stationary limit of the mean square displacement $s(t)$  
determined by the ratio of the diffusive constant versus the memory strength: $s(t \to \infty) \propto D/\lambda$. 
The reason for such a new behavior is by means of an explicit coupling of 
the rate of the concentration at the observation time $t$ to that one at a previous time. This time accumulation 
is further accompanied by an additional spatial accumulation, the effect of which is comparable to the effect 
a long-range interaction forces and consequently the results are basically independent on the spatial  
dimensions in according to scaling arguments. These many-body effects are shown to change the asymptotic 
behavior drastically. Due to the feedback-coupling of a particle to its environment, a subsequent particle, 
undergoing a diffusive motion, gains information from a modified environment. Hence the particle can be 
confined within a certain region preventing a further transport. In this manner a self-organized memory leads 
to a non-zero stationary mean square displacement controlled by the memory strength.\\ 
A further application is given by a non-linear diffusive process where the memory is also originated by gradient 
terms with the strength $\mu $. Although the additional non-linear memory term offers the same scaling behavior 
as conventional diffusion the resulting probability distribution is completely different from diffusion. 
Whereas the second moment, the mean square displacement, shows a diffusive behavior, higher order cumulants, in 
particular Binder`s cumulant
$$
C_B(t) = 1- \frac{w(t)}{3 s^2(t)} = \frac{2}{3d}\left[1-d+\frac{\mu}{4D}(d+4)\right]
$$  
reveals deviations from conventional diffusion. Especially, $C_B$ is determined by the strength of the 
memory. The analysis will be extended to chemical reactions in multicomponent systems.

\begin{acknowledgments} 
This work was supported by the DFG (SFB 418).
\end{acknowledgments}

\end{document}